\documentclass[aps,pra,twocolumn,superscriptaddress,showkeys]{revtex4-2}

\usepackage{amsmath, amssymb, mathtools, physics, dsfont, hyperref, orcidlink, microtype, graphicx}
\usepackage[T1]{fontenc}
\usepackage[dvipsnames]{xcolor}
\usepackage{tikz} %quantum circuits
\usetikzlibrary{quantikz}

\hypersetup{
	colorlinks=true,
	linkcolor=Brown,
	citecolor=Blue,
	filecolor=magenta,
	urlcolor=RoyalPurple
}

\begin{document}

\title{Quantum Vault: Symmetric No-Cloning Protection for Token Issuer and User (Benchmarked on IBMQ)}

\thanks{This version supersedes v1 (\href{https://arxiv.org/abs/2605.03564}{arXiv:2605.03564v1}) posted in arXiv, which contained a data analysis error in the averaging of the shots within some of the Qiskit jobs, and a misconception regarding state collapse upon measurement of the auxiliary qubit in the SWAP test.
	The protocol has been re-implemented and the results re-analyzed accordingly with these corrections. }

\author{Lucas Tsunaki \orcidlink{0009-0003-3534-6300}}
\email{tsunakilucas@gmail.com}
\affiliation{Department Spins in Energy Conversion and Quantum Information Science (ASPIN), Helmholtz-Zentrum Berlin für Materialien und Energie GmbH, Hahn-Meitner-Platz 1, 14109 Berlin, Germany}

\author{Boris Naydenov \orcidlink{0000-0002-5215-3880}}
\email{boris.naydenov@helmholtz-berlin.de}
\affiliation{Department Spins in Energy Conversion and Quantum Information Science (ASPIN), Helmholtz-Zentrum Berlin für Materialien und Energie GmbH, Hahn-Meitner-Platz 1, 14109 Berlin, Germany}
\affiliation{Department of Physics, Freie Universität Berlin, Arnimallee 14, 14195 Berlin, Germany}

\date{\today}

\begin{abstract} 
Quantum tokens are underlying primitives for quantum money and network proposals, which leverage the no-cloning theorem to realize unforgeable authentication. 
A relevant but overlooked type of attack to such architectures is one in which a hacker steals the classical side information about the token states from the issuing agent (e.g., a bank), enabling the forgery of fake tokens without violating the no-cloning theorem.
Our proposal addresses this threat by removing the classical side information about the token states, where instead a copy of the token is stored at the bank, i.e. a quantum vault.
This copy can be accessed by anyone to perform authentication, consuming the token pair in the process.
We benchmark the protocol and identify quality parameters within a hardware agnostic framework employing IBMQ processors, such that the protocol is applicable to arbitrary quantum platforms.
By comparing the efficiency with which genuine tokens are produced and authenticated against a possible estimate-and-forge attack, we quantify the security of the protocol.
We achieve probabilities lower than $10^{-4}$ for false-negative errors (Type 1) and $10^{-10}$ for successful attacks (Type 2) when considering quantum bills composed of 200 tokens, even in the worst performing processor and most conservative uncertainty estimate.
The quantum vault not only protects both user and bank symmetrically with the same quantum principles, but provides a step towards public key authentication, since any untrusted party can have access granted by the bank to the tokens' authentication interface without being able to clone them.
Besides this bank-mediated form of publicly accessible verifiability, our proposal achieves traceability and revocability, since the bank retains the vault copy of the tokens.
\end{abstract}


\maketitle

%%%%%%%%%%%%%%%%%%%%%%%%%%%%%%%%%%%%%%%%%%%%%%%%%%%%%%%%%%%%%%%%%%%%%%%

\section{Introduction}\label{sec:introduction}

\begin{figure*}[t!]
	\includegraphics[width=\textwidth]{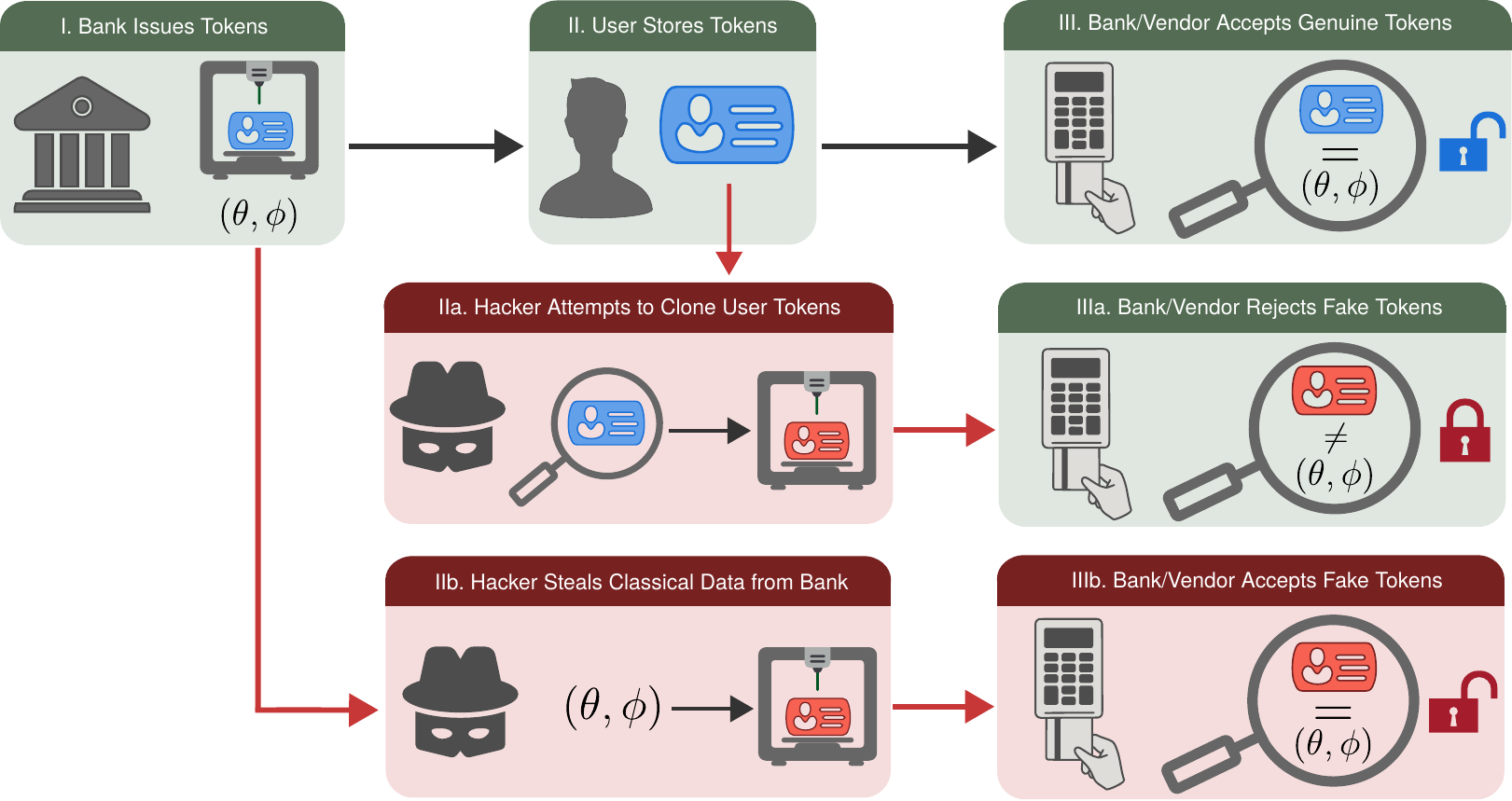}
	\caption[Attack Scenarios on Previous Quantum Token Proposals with Classical Side Information]{Attack scenarios on previous quantum token proposals with classical side information.
	In general, a token protocol has three steps: \textbf{I.} a bank issues quantum tokens in states $\ket{\psi}$ with angles $(\theta, \phi)$; \textbf{II.} the bank transmits the tokens to the user, who stores them until authentication; \textbf{III.} a vendor or bank authenticates the tokens by measuring their states and comparing them with the classically stored values of $(\theta, \phi)$.
	While the user's token is protected by the no-cloning theorem, an attacker could steal the values of $(\theta, \phi)$ from the bank and create fake tokens without the bank realizing that it had been compromised.}
	\label{fig:attack_scheme}
\end{figure*}

Quantum communication infrastructure and technology are becoming central components for national sovereignty and geopolitics~\cite{qgeopolitics, qkd_china_1, qcoridor, qunet, demoquandt}, with robust applications~\cite{qcomm_review, qkd_china_2, idquantique} achievable within current NISQ-era~\cite{nisq2fasq} hardware.
The foundation for quantum communication lies in the no-cloning theorem, which states that arbitrary quantum states cannot be cloned in general~\cite{no_cloning_1, no_cloning_2, no_cloning_3, encrypted_qubits_cloning}.
While most applications rely on entanglement and photon exchange to communicate information, another paradigm within quantum communication is a physical object which holds a non-clonable quantum state that can be used for personal authentication, i.e. a quantum token.
This concept was originally proposed by S. Wiesner in 1983~\cite{qtoken_original} and since then, overlapping applications have been introduced under different names: quantum money~\cite{qtoken_original, qtoken_attack3}, quantum ticket~\cite{quantum_token}, quantum payment~\cite{qpayment}, quantum signature~\cite{qsignature}, quantum physical unclonable function~\cite{review_qpuf}, quantum identity authentication~\cite{qia} or quantum token~\cite{qtoken_1, qtoken_2, qtoken_tim, grand_challenge}.
Meanwhile, the same terms often represent slightly distinct concepts across literature~\cite{review_qpuf}.

To be precise, we consider in this work a personal quantum token with classical state information as an application consisting of three steps (Fig.~\ref{fig:attack_scheme}):
\begin{enumerate}
	\item \textbf{Token Issuing:} an issuing agent (e.g. a bank or mint) prepares an arbitrary quantum state $\ket{\psi}$ with Bloch sphere angles $(\theta, \phi)$ known only to itself and classically stores this information about the prepared state.
	\item \textbf{Token Storage:} the issuing agent remotely transmits or physically delivers the state $\ket{\psi}$ to the user, or claimant~\cite{nist}, who stores it until authentication is required.
	\item \textbf{Token Authentication:} the user presents their token to the authentication agent referred to as verifier~\cite{nist}, which can be a vendor, the bank or another user.
	The verifier measures this state and compares with the initially prepared angles $(\theta, \phi)$ and if the measured state is close up to some acceptance range to the prepared state, the token is accepted.
\end{enumerate}
Here we use a terminology closer to quantum money applications, but in fact quantum tokens have broader applications for general authentication procedures.
An important distinction from other quantum money proposals~\cite{anonymous_qtoken, public_qcoins} is that, here, we consider personal non-anonymous tokens, in analogy to classical credit cards compared to cash.
This common quantum token definition presupposes secure classical communication channels between parties~\cite{semiq_money, qtoken_attack4,  public_qcoins, public_key_qmoney} to communicate $(\theta, \phi)$ and/or classical storage in the bank~\cite{qtoken_original, quantum_token}, which is precisely the vulnerability addressed in this work.
This definition also carries a classical side information about the serial number of the token for user identification, which still remains in our proposal.

In simplified terms, a quantum token can thus be understood as an unclonable quantum memory equipped with a secure readout method.
In a public-key architecture, any individual is able to verify these tokens using the readout method, but not clone them~\cite{classical_bank, public_key_qmoney, public_qcoins}.
Conversely, in a private-key setting, only the bank or a trusted party such as a vendor is able to authenticate the token.
As will be shown here, our proposal trivially works as a private-key architecture and can operate with public-accessible verifiability mediated by the bank, if there exists a quantum channel between the bank and the verifier.

Quantum tokens can either be thought of as a portable object which the bank physically gives to the users (analogous to classical credit cards), forming a fundamental component for quantum money schemes~\cite{qmoney_1, qtoken_attack3}, or as a more abstract concept, where the token state is transmitted through flying qubits or teleported and stored locally by users, within quantum internet proposals~\cite{qinternet_1, qinternet_2, semiq_money} and other quantum network applications~\cite{qtoken_network_authentication}.
In any case, the implementation of such a device is still a long way from application-scale and commercial stages of development due to various practical constraints~\cite{token_fab_1, token_fab_2}.

\begin{figure*}[!t]
	\includegraphics[width=\textwidth]{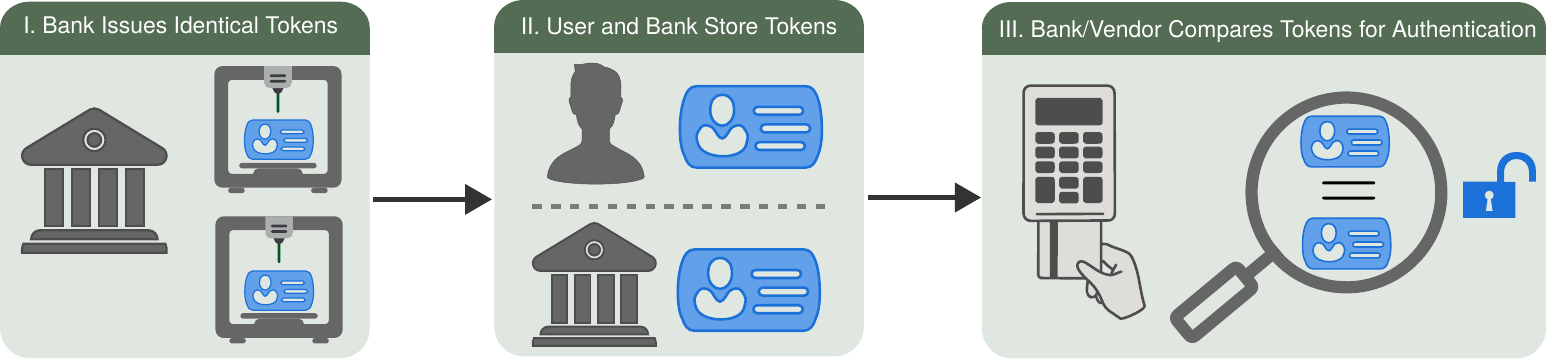}
	
	\begin{quantikz}[font=\large]
		\lstick{$\ket{0}^{\otimes N}$} & \gate{\hat{R}(\theta, \phi)} & \rstick{$\ket{\psi}^{\otimes N}$} \qw \\
		\lstick{$\ket{0}^{\otimes N}$} & \gate{\hat{R}(\theta^\prime, \phi^\prime)} & \rstick{$\ket{\psi^\prime}^{\otimes N}$} \qw
	\end{quantikz}
	{\huge $\Longrightarrow$}\hspace{.6cm}
	\begin{quantikz}[font=\large]
		\lstick{$\ket{\psi}^{\otimes N}$}
		\\ \\ \\
		\lstick{$\ket{\psi^\prime}^{\otimes N}$}
	\end{quantikz}
	\hspace{1cm}{\huge $\Longrightarrow$}\hspace{1cm}
	\begin{quantikz}[font=\large]
		\lstick{$\ket{\psi}^{\otimes N}$} & \qw & \targX{} & \qw & \qw \\
		\lstick{$\ket{\psi^\prime}^{\otimes N}$} & \qw & \targX{} & \qw & \qw  \\
		\lstick{$\ket{0}^{\otimes N}$} & \gate{H} & \ctrl{-2} & \gate{H} & \meter{} \vcw{1} \\
		\lstick{$c_n$} & \cw & \cw & \cw & \cwbend{-1}
	\end{quantikz}
	\caption[Quantum Vault Protocol]{
		Quantum vault protocol.
		Initially, the bank issues two tokens in identical Haar random states $\ket{\psi}^{\otimes N}=\ket{\psi^\prime}^{\otimes N}$ and immediately discards any classical information about them.
		Each token is an ensemble of $N$ non-entangled independent qubits.
		One token is stored in the bank's vault while the other is given to the user, such that both parties are no-cloning protected from attacks.
		For authentication, the bank or vendor performs SWAP tests between the $N$ qubits of the tokens to verify if both tokens are in the same state or if one of them is fake.}
	\label{fig:qvault_protocol}
\end{figure*}

Beyond the discussion around practical feasibility, a malicious attacker would not be able to efficiently clone a token held by the user~\cite{qtoken_attack, qtoken_attack2, qtoken_attack3, qtoken_attack4} (case IIa in Fig.~\ref{fig:attack_scheme}).
Within an estimate-and-forge model~\cite{review_qpuf}, this attacker can try to measure the user's token to gain some information about the state $\ket{\psi}$ and make a fake token with this information.
However, as recently demonstrated within IBM Quantum platforms (IBMQ), the bank can distinguish genuine tokens from forged ones even in the presence of experimental errors~\cite{qtoken_1, qtoken_2}.
Another known type of attack~\cite{qmoney_3, qmoney_4}, but rarely covered by quantum token proposals~\cite{grand_challenge}, is a hacker that attacks the bank instead of the user.
In most applications, the issuing agent is assumed to be an idealized entity which cannot be hacked~\cite{bsi_qkd_attacks}.
But if an attacker is able to steal or intercept the classical information of the quantum token $(\theta, \phi)$ (case IIb in Fig.~\ref{fig:attack_scheme}), they would be able to create copies of the genuine token without violating the no-cloning theorem and the verifier would accept such tokens without even realizing the security breach (case IIIb in Fig.~\ref{fig:attack_scheme}).
Such a scenario might have been considered unrealistic until a few years ago, but recent events such as the XZ Utils backdoor in 2024~\cite{backdoor}, the multi-billion-dollar North Korean crypto heists in 2025~\cite{cryptoheist} or the Axios attack in 2026~\cite{axios_hack} have highlighted security vulnerabilities in the digital infrastructure and databases from financial institutions, as well as from other sectors.
Therefore, a quantum verification scheme that also protects the bank is not simply a minor security improvement to previous proposals, but a crucial requirement for future quantum money and quantum internet architectures.

To answer this previously overlooked question on how to implement a protocol that protects both parties symmetrically with the same fundamental quantum principles, we introduce the concept of the quantum vault, as shown in Fig.~\ref{fig:qvault_protocol} (the name quantum vault was already introduced in Ref.~\cite{qvault}, but in a different context).
Simply put, instead of preparing a quantum token and storing the information about the angles $(\theta, \phi)$ classically as in prior proposals, the issuing agent now prepares two quantum tokens in the same state $\ket{\psi}^{\otimes N}$ and gives one of them to the user, while the other is stored, i.e. the quantum vault.
We consider ensemble-based quantum tokens~\cite{qtoken_1, qtoken_2} composed of $N$ non-entangled independent qubits, in order to improve their readout efficiency. 
The authentication can then be performed by any party, trusted or not, by comparing the user's token with the one stored in the quantum vault.
By eliminating the classical side information about $(\theta, \phi)$, we are not only able to achieve public-accessible verifiability mediated by the bank, but we also demonstrate unforgeability, traceability and revocability powers.

Before a scheme like this can be implemented in practical large-scale applications with arbitrary systems, the protocol needs to be benchmarked and thoroughly analyzed within a hardware agnostic framework.
By doing so, key parameters can be identified and the security of the proposed scheme against attacks can be quantified.
To realize that, we employ three IBMQ processors using Qiskit SDK~\cite{qiskit}, where the implementations are open-source provided at a GitHub repository~\cite{git_repo}.
More details are given in Sec.~\ref{sec:methods}, while Table~\ref{tab:variables} provides a summary of all mathematical variables introduced in the text.
We begin this work in Sec.~\ref{sec:protocol} developing the quantum vault protocol and benchmarking it on the IBMQ hardware.
Following on that, Sec.~\ref{sec:security} implements an estimate-and-forge attack, quantifying the security of the protocol.
Finally, in Sec.~\ref{sec:discussion}, we discuss limitations and possible improvements for the protocol, outlining how the protocol could be implemented in practice.
\section{Quantum Vault Protocol}\label{sec:protocol}

\begin{figure*}[t!]
	\centering	
	\includegraphics[width=\textwidth]{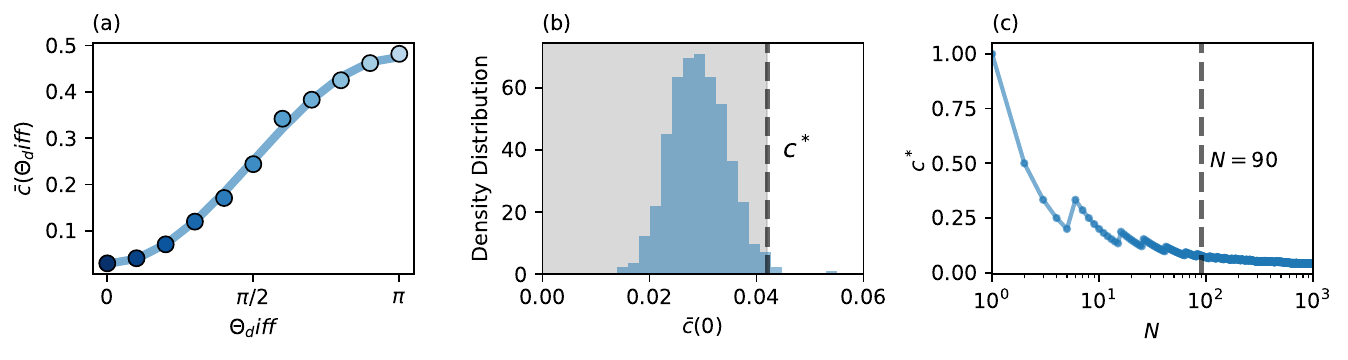}
	\caption[Benchmark of quantum vault protocol in IBMQ Kingston]{
		Benchmark of quantum vault protocol in IBMQ Kingston.		
		\textbf{(a)} SWAP test observable $\bar{c}(\Theta)$ for ensemble size $N=1000$ (error bars are smaller than points).
		At $\Theta=0$, $\bar{c}$ is close to 0, which then increases with $\Theta$, enabling the verifier to authenticate if $\ket{\psi}^{\otimes N} = \ket{\psi^\prime}^{\otimes N}$.
		The quality parameters $Q_a$ and $Q_o$ are obtained from the fit of the curve.
		\textbf{(b)} Distribution of $\bar{c}(0)$ from 400 angles $(\theta, \phi) = (\theta^\prime, \phi^\prime)$ uniformly distributed on the Bloch sphere with $N=1000$.
		The acceptance threshold $c^*$ is defined such that the probability that the bank accepts genuine tokens is $p_b[\bar{c} \leq c^*]=0.99$, corresponding to the shaded area of the distribution below $c^*$.
		\textbf{(c)} Acceptance threshold $c^*$ as a function of ensemble size.
		As $N$ increases, the shot noise of the distribution decreases, leading to a more stringent acceptance criterion represented by the decrease in $c^*$.
		The ensemble size is taken as $N=90$, such that $c^*$ is not saturated and the discretization noise is already low.
	}
	\label{fig:benchmark}
\end{figure*}

In our proposal for the quantum vault scheme (Fig.~\ref{fig:qvault_protocol}), the bank starts by preparing two tokens in identical Haar random states $\ket{\psi}^{\otimes N}=\ket{\psi^\prime}^{\otimes N}$ through a rotation applied to the $2N$ qubits as $\ket{\psi}^{\otimes N} = \hat{R}^{\otimes N}(\theta, \phi) \ket{0}^{\otimes N}$ and $\ket{\psi^\prime}^{\otimes N} = \hat{R}^{\otimes N}(\theta^\prime, \phi^\prime) \ket{0}^{\otimes N}$, which does not violate the no-cloning theorem.
The angles $(\theta, \phi)$ are randomly drawn on the Bloch sphere, which could be done through a quantum random number generator~\cite{qrng}, thus potentially reducing risks from side-channel attacks during token preparation~\cite{review_qpuf, side_channel_attack, bsi_qkd_attacks}.
During storage, both parties can use error correction and mitigation techniques~\cite{qe_mitigation, review_qpuf} to avoid decoherence of the states, such as dynamical decoupling from noises, as implemented in similar experiments with IBMQ~\cite{dd_ibmq}.

Finally, during authentication, the user presents their token to the verifier for comparison with the one stored by bank in the vault.
In a private-key setting, where the bank is both the issuing and verifier, this comparison is direct, since the bank holds both the token stored in the vault and the one presented by the user.
In the publicly accessible verification architecture on the other hand, the bank needs to allow anyone to verify the validity of the token, being a trusted vendor or even another untrusted user.
For that to be realized, the verifier requests the bank to start an authentication procedure, where the bank can transfer the token stored in the vault to the verifier or allow the verifier to perform the authentication in the vault.
This can be done through teleportation~\cite{teleportation, teleport2} or flying qubits conversion~\cite{flying_qubit, grand_challenge, qtoken_tim}, both of which require a quantum channel between the bank and the verifier based on existing quantum network technology~\cite{qkd_china_1, qcoridor, qunet, demoquandt}.
Arguably, teleportation via heralded entanglement is a more viable alternative, since the tokens are safely stored in the vault until an entanglement is heralded, while flying qubits irretrievably loses many qubits. 
In connection to another wide-spread technology, a similar procedure takes place when a card payment is made by a user to a vendor, mediated by a centralized payment processor.
The vendor reads the user's card information and through a classical channel with the payment processor, verifies the veracity of their token.
However, in our proposal for the quantum vault, the quantum channel does not necessarily need to be secure and the vendor does not have to be trusted, as demonstrated in Sec.~\ref{sec:discussion}.

The main technical question is then what is the most efficient way for the verifier to authenticate if both states are the same or if they have been cloned.
S. Barnett et al. (2003)~\cite{barnett2003comparison} theoretically demonstrated that it is impossible to identify if two single pure unknown states are identical with absolute certainty and established certain conditions for a measurement that compares two unknown quantum states to be optimal.
This motivates the adoption of ensembles of independent qubits as tokens to improve the states comparison measurement.
The use of ensembles as quantum tokens can further facilitate the technological implementation of such technologies in certain physical systems~\cite{token_fab_1, grand_challenge}.
However, the no-cloning theorem is not trivially applicable in these cases, given that an ensemble is already a copy of individual qubits and an attacker can exploit that to perform quantum tomography of the $\ket{\psi}^{\otimes N}$ states, thus achieving a better estimate of $\ket{\psi}$.
Refs.~\cite{qtoken_1, qtoken_2} have nevertheless shown that ensemble-based quantum token protocols are still viable, even when the attacker has access to better quantum tomography capabilities.

One possible method which satisfies the optimal conditions for quantum state comparison~\cite{barnett2003comparison} is the SWAP test~\cite{swap_test1, swap_test2, swap_test_review}, originally introduced for quantum fingerprinting applications, but which has also been proposed to be used for quantum token authentication~\cite{qtoken_network_authentication, public_qcoins}.
In the SWAP test, an auxiliary qubit is added to the system, as shown in Fig.~\ref{fig:qvault_protocol}.
First, a Hadamard gate is applied to the auxiliary qubit, followed by a controlled SWAP between the token qubits and another Hadamard to the auxiliary.
Lastly, the auxiliary qubit is measured in the standard basis $\{ \ket{0}, \ket{1}\}$ and the measurement outcome is stored in the classical register $c_n$ of the $n$-th test, with values 0 or 1.
We define the SWAP test observable as the average of the classical register over the $N$ qubits,
\begin{equation}
	\bar{c} = \frac{1}{N} \sum_{n=1}^{N} c_n .
\end{equation}
While $c_n$ can only be 0 and 1, $\bar{c}$ has $N+1$ possible discrete values between 0 and 1, converging to a continuous variable for large $N$.
The value of $\bar{c}$ depends on the projection between the two states as~\cite{swap_test1, swap_test2}
\begin{equation}\label{eq:Pc1}
	\bar{c} = \frac{1 - | \bra{\psi} \ket{\psi^\prime} |^2}{2} .
\end{equation}
This way, without the presence of errors, the measurement outcome is $\bar{c}=0$ if both states are the same $\ket{\psi}^{\otimes N} = \ket{\psi^\prime}^{\otimes N}$~\cite{swap_test_review} and $\bar{c}>0$ if they are not identical.
In this sense, a single SWAP test is not conclusive, since a single measurement $c_n=1$ determines that both states are not perfectly the same, but $c_n=0$ can occur for both if the states are the same or not, therefore justifying the adoption of ensembles of qubits as tokens instead of single ones.

We can define a separation angle $\Theta$ between $\ket{\psi}$ and $\ket{\psi^\prime}$ given by $\Theta = \arccos (\hat{n} \cdot \hat{n}^\prime)$, where $\hat{n}$ and $\hat{n}^\prime$ are the unit vectors along the states $\ket{\psi}$ and $\ket{\psi^\prime}$ over the Bloch sphere.
After measuring $N$ independent SWAP tests between $\ket{\psi}^{\otimes N}$ and $\ket{\psi^\prime}^{\otimes N}$, the average value $\bar{c}$ from Eq.~\ref{eq:Pc1} in terms of $\Theta$ is
\begin{equation}\label{eq:c_theta}
	\bar{c}(\Theta) = Q_{o} + \frac{Q_{a}}{4} \left[1 - \cos( \Theta) \right] .
\end{equation}
$Q_{a}$ and $Q_{o}$ are defined in such a way that they take into account experimental imperfections in the implementation, therefore representing quality parameters for the quantum vault hardware.
$Q_{a}$ corresponds to the oscillation amplitude or contrast between identical and orthogonal states $Q_{a} = 2[\bar{c}(\pi) - \bar{c}(0)]$, ideally being equal to 1 in the absence of experimental errors.
While $Q_{o}$ corresponds to an offset of the oscillation with $Q_{o} = \bar{c}(0)$, ideally being 0.

\begin{figure*}[t!]
	\centering	
	\includegraphics[width=\textwidth]{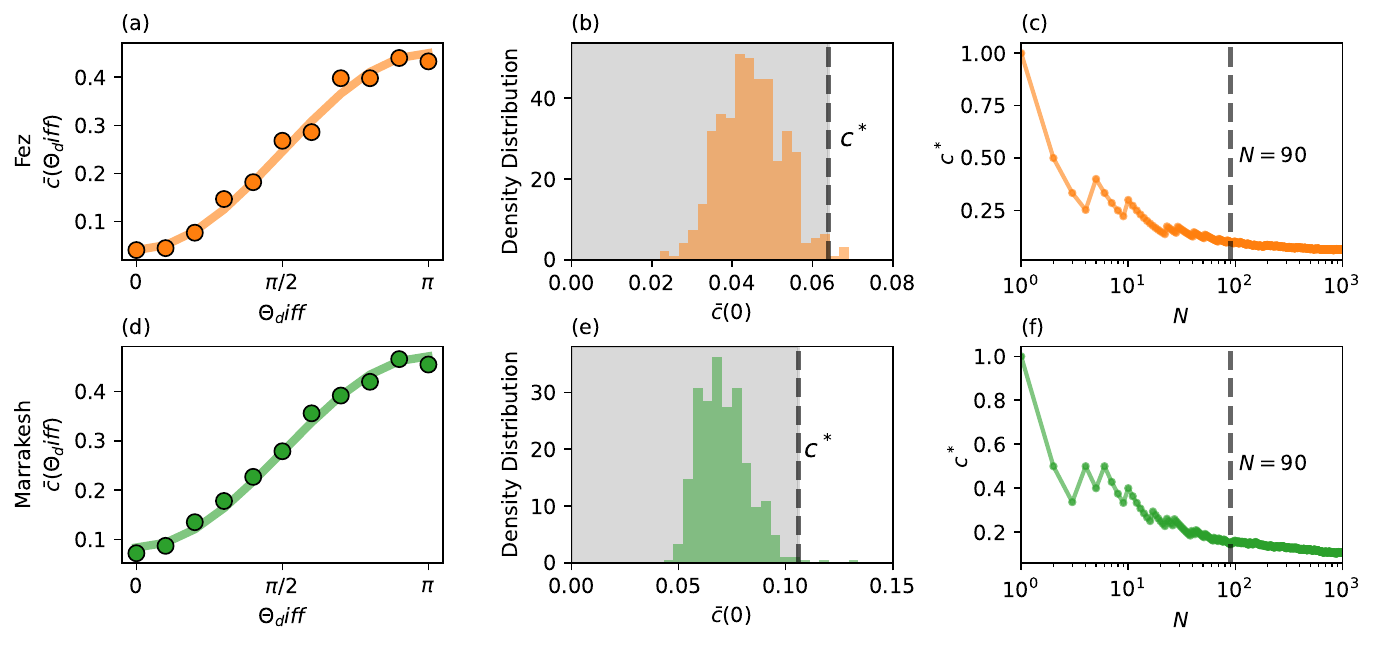}
	\caption[Quantum Vault Benchmark in IBMQ Fez and Marrakesh]{
		Quantum Vault Benchmark and Attack in IBMQ Fez and Marrakesh.
		\textbf{(a, d)} SWAP test observable $\bar{c}$ as a function of separation angle $\Theta$.
		\textbf{(b, e)} $\bar{c}(0)$ distributions for 400 Bloch sphere points.
		\textbf{(c, f)} Acceptance threshold $c^*$ of the SWAP test as a function of the ensemble size $N$.
	}
	\label{fig:sup_fig}
\end{figure*}

In order to obtain these quality factors and characterize the SWAP test response to different separations angles, $\bar{c}(\Theta)$ was measured between $\ket{\psi}^{\otimes N}=\ket{0}^{\otimes N}$ and $\ket{\psi^\prime}^{\otimes N}=\hat{R}^{\otimes N}(\Theta, 0)\ket{0}^{\otimes N}$ with $N=1000$, as shown in Fig.~\ref{fig:benchmark}~(a) for IBMQ Kingston and in Fig.~\ref{fig:sup_fig}~(a, d) for IBMQ Fez and Marrakesh.
We assume an ergodic approximation for the ensemble, where the $N$ qubits are time averaged in independent sequential circuit shots instead of physical realizations of distinct qubits, thus reducing computational costs.
When the two states $\ket{\psi}$ and $\ket{\psi^\prime}$ are close with $\Theta=0$, the average of the SWAP test is close to 0, represented by $Q_o$.
But as $\Theta$ increases and the states become orthogonal, $\bar{c}(\Theta)$ gets closer to 0.5.
At this value of $N$, the error bars given by the standard deviation of the measurement are smaller than the points, being dominated by shot noise~\cite{qtoken_1}, as can be visualized in the GitHub repository~\cite{git_repo}.
By fitting the data with Eq.~\ref{eq:c_theta}, we obtain $Q_{o}$ and $Q_{a}$, as given in Tab.~\ref{tab:params}.
Kingston has better quality parameters than Marrakesh and Fez, given its lower readout and gate errors, together with longer coherence times.

Based on the outcome of the SWAP tests, the bank needs to determine an acceptance criterion for the tokens.
With that, we define an acceptance threshold $c^*$ such that any measurement of $\bar{c}$ below or equal to $c^*$ is considered as accepted.
This acceptance threshold is a free parameter by construction that can be adjusted by the bank to achieve desired metrics of security.
Here, we fix $c^*$ such that the probability for accepting genuine tokens is 
\begin{equation}\label{eq:pb_c*}
	p_b[\bar{c} \leq c^*] = 0.99.
\end{equation}
To obtain a distribution for $\bar{c}(0)$ and thus calculate the $c^*$ values, we take 400 points $(\theta, \phi)=(\theta^\prime, \phi^\prime)$ uniformly distributed on the Bloch sphere with $N=1000$ in the three processors, as shown in Fig.~\ref{fig:benchmark}~(b) for IBMQ Kingston and Fig.~\ref{fig:sup_fig}~(b, e) for the other two hardware.
Comparatively to each other, Kingston has a $\bar{c}(0)$ distribution concentrated at smaller values, where the shaded area below $c^*$ corresponds to the acceptance probability $p_b=0.99$.

The corresponding values of $c^*$ obtained from the percentiles from Eq.~\ref{eq:pb_c*} as a function of $N$ are shown in Fig.~\ref{fig:benchmark}~(c) and Fig.~\ref{fig:sup_fig}~(c, f).
As $N$ increases, the shot noise of the distribution decreases, leading to a better estimate of $\bar{c}$ with smaller width and, consequently, lower acceptance threshold $c^*$.
A lower acceptance threshold represents a more stringent acceptance criterion, which directly translates to a higher security of the protocol.
As $N$ increases beyond $10^2$ however, $c^*$ saturates, such that adding more qubits in the ensemble does not significantly decrease $c^*$.
Another important observation is that $c^*$ does not decrease monotonically, due to the discrete nature of the $\bar{c}$ variable and the finite sampling of the Bloch sphere.
A sensible choice for the ensemble size is therefore $N=90$, such that the $c^*$ has not yet reached the saturated regime, but $\bar{c}$ has already enough possible values such that the discretization noise is low.
This ensemble size is also close to the $N=100$ used in Ref.~\cite{qtoken_1} for the implementation of ensemble-based quantum token protocols in IBMQ processors, but here we also adopt a value divisible by 3 in order to simplify the security analysis (Sec.~\ref{sec:security}).
The corresponding values of $c^*$ at $N=90$ are given in Tab.~\ref{tab:params}.

\begin{figure}[b!]
	\includegraphics[width=\columnwidth]{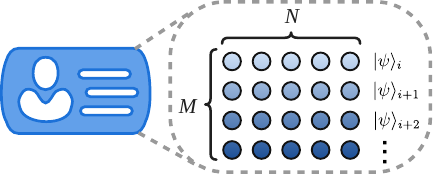}
	\caption[
	Proposed Quantum Bill Architecture
	]{
		Proposed quantum bill architecture in the state $\ket{\mathbf{\Psi}} = \bigotimes^M_{m=1} \ket{\psi}_m^{\otimes N}$.
		$M$ quantum token states are encoded in the bill, each with $N$ qubits.
	}
	\label{fig:qbill}
\end{figure}

Another safety feature of the quantum vault method comes from the fact that quantum token authentication devices are typically composed of several tokens, also referred to as quantum coin~\cite{multicopies, public_qcoins}, banknote~\cite{qtoken_attack, qtoken_attack3} or bill~\cite{public_qcoins}.
Since `coin' is usually used for anonymous untraceable tokens, we use the quantum bill terminology.
The verifier thus needs to test if $\ket{\psi}_m^{\otimes N} = \ket{\psi^\prime}_m^{\otimes N}$ for various states $m=1,...,M$.
This proposed architecture is represented in Fig.~\ref{fig:qbill}, with a total of $M$ independent tokens, each with $N$ qubits.
The quantum state of the whole quantum bill thus corresponds to
\begin{equation}\label{eq:Psi}
	\ket{\mathbf{\Psi}} = \bigotimes^M_{m=1} \ket{\psi}_m^{\otimes N} .
\end{equation}

\begin{figure}[b!]
	\centering	
	\includegraphics[width=\columnwidth]{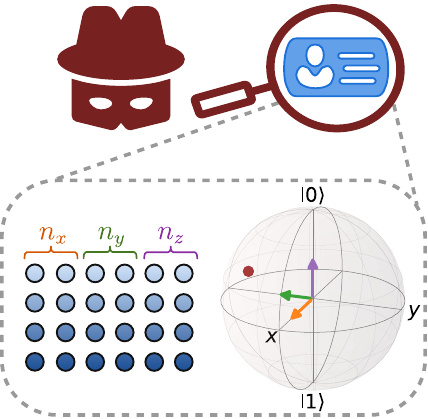}
	\caption[Estimate-and-Forge Attack Method]{
		Estimate-and-Forge attack method.
		The attacker subdivides the ensemble in the $\ket{\psi}^{\otimes N}$ state into three equal parts to perform SWAP test measurements along the $x$, $y$ and $z$ axes.
		With this information, they use maximum-likelihood estimation to reconstruct $\ket{\psi}$ and then forge fake tokens and attempt authentication with the bank. 
	}
	\label{fig:attack_bloch}
\end{figure}

Given that the probability that a single token is accepted by the bank is $p_b$, the probability $\mathcal{P}_b$ that at least $m^*$ token states out of a total $M$ are accepted corresponds to a cumulative binomial distribution as
\begin{equation}\label{eq:Pb}
	\mathcal{P}_b(m^*, M, p_b) = \sum_{m=m^*}^M \binom{M}{m} p_b^{m} (1-p_b)^{M-m} .
\end{equation}
Likewise, $m^*$ acts as another threshold parameter for the minimum number of token states which need to be accepted in the bill.
By adjusting the values of $m^*$ and $M$, desired levels for $\mathcal{P}_b$ can be achieved as well.
If the acceptance threshold $m^*$ is lowered, the Type 1 error rate of rejecting genuine tokens can be decreased.
But on the other hand, a lower threshold $m^*$ leads to higher chances of Type 2 errors for accepting fake tokens, as further discussed in Sec.~\ref{sec:security}.
In this work, we choose $m^*$ such that $\mathcal{P}_b \geq 1-10^{-4}$.
In addition, increasing the total number of states in the bill $M$ leads to a more stringent acceptance threshold $m^*$, again improving the overall security~\cite{multicopies, qtoken_1, qtoken_2}.

Prior to the application of the quantum vault protocol, the bank thus needs to perform a calibration procedure as developed here to obtain the $\bar{c}(0)$ distribution and the acceptance threshold $c^*$.
By doing so, the measurement of the variable $\bar{c}$ by the SWAP test can be used to verify if the two bills are in the same states $\ket{\mathbf{\Psi}}$ and consequently authenticate the identity of the user.
The only classical information involved in the protocol is therefore $c^*$ and the serial number to identify the tokens, completely eliminating any information about the angles $(\theta, \phi)_m$.
These classical parameters can even be made public for a public accessible authentication architecture.
The values of $Q_a$ and $Q_o$ are not strictly necessary for the practical implementation of the method, but they can still be valid quality parameters for comparing different hardware.
%------------------------------------------------------

\section{Security Against Estimate-and-Forge Attack}\label{sec:security}

Having benchmarked the quantum vault protocol and characterized the efficiency in which tokens are issued and authenticated, we now implement a possible estimate-and-forge attack scenario.
By comparing the efficiency which an attacker can forge fake tokens with the efficiency in which the bank can generate and authenticate genuine ones, we can quantify the security of the protocol.
The quantum vault protocol requires that the hardware can execute a SWAP test circuit.
A fair assumption is that an attacker could have access to the same authentication interface to realize SWAP tests.
Moreover, in a publicly accessible authentication architecture, untrusted parties already need to interact with the vault's tokens through the SWAP test authentication interface anyway.
We thus consider an attack model where the attacker attempts to gain information about the token states and then creates fake ones for authentication with the verifier (case IIa in Fig.~\ref{fig:attack_scheme}).
Both user and bank are symmetrically protected by the no-cloning theorem, but if the attacker attempts to hack the vault's tokens, then no authentication can be performed afterwards, since the vault's copy is destroyed in the process.
We therefore consider the case where the attacker attempts to hack the user's token, even though this security analysis is valid for both token holders.

\begin{figure*}[t!]
	\centering	
	\includegraphics[width=\textwidth]{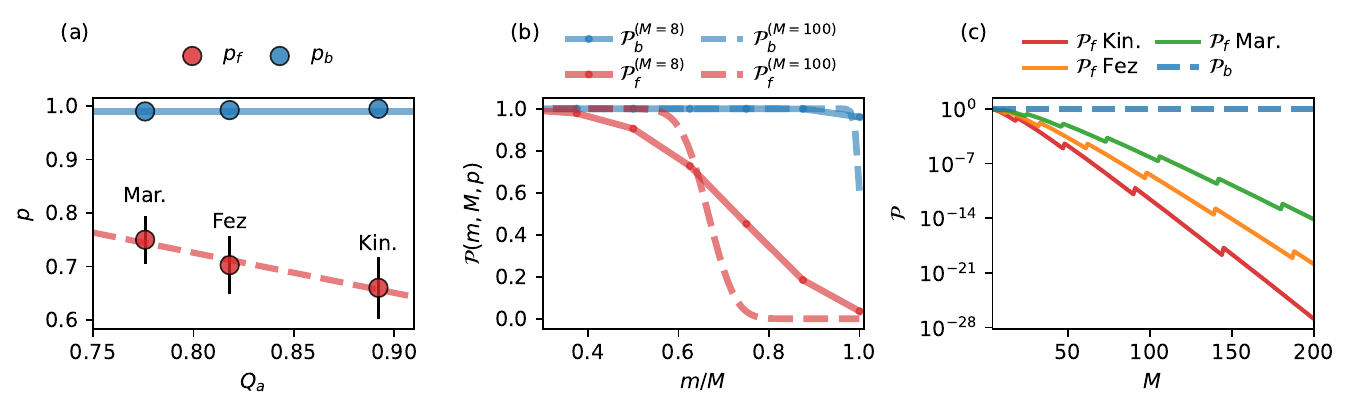}
	\caption[Forgery Attack on Quantum Vault]{
		Forgery Attack on Quantum Vault.
		\textbf{(a)} Acceptance probabilities $p_b$ for single genuine tokens and forged ones $p_f$, with error bars obtained from bootstrapping~\cite{bootstrap}.
		$p_f$ is constant within error bars, but agrees with a linear decrease with $Q_a$ (red guide line), as observed in Refs.~\cite{qtoken_1, qtoken_2}.
		\textbf{(b)} Probability to accept genuine and forged bills $\mathcal{P}(m^*, M, p)$ composed of $M=8,\; 100$ tokens as a function of $m^*/M$ in IBMQ Kingston.
		Both probabilities decrease as the acceptance threshold $m^*/M$ increases, but faster with $\mathcal{P}_f$.
		With larger $M$, the cumulative binomial distribution approaches a rectangular function, improving security.
		\textbf{(c)} Probability of successful forgery $\mathcal{P}_f(M, p)$ for $m^*$ fixed such that $\mathcal{P}_b\geq1-10^{-4}$ in all three IBMQ.
		$\mathcal{P}_f$ decays exponentially with $M$, reaching negligible values at large $M$, even in the worst performing hardware.
	}
	\label{fig:attack_benchmark}
\end{figure*}

In order to gain information about the genuine token states $\ket{\psi}^{\otimes N}$ along the unit vector $\hat{n} = (n_x, n_y, n_z)= (\sin\theta \cos\phi, \sin\theta \sin\phi, \cos\theta)$, the attacker can prepare known probing states and perform SWAP tests to estimate the angle $\Theta$ between them.
By setting this probing state as $\ket{+X}$, the attacker measures $\bar{c}_x$ which can be used to estimate the $n_x$ component of $\ket{\psi}$.
Conversely, with $\ket{+Y}$ and $\ket{0}$, they obtain $\bar{c}_y$ and $\bar{c}_z$, which can be used to estimate the $n_y$ and $n_z$ components respectively.
From the measured values of $\bar{c}_{x,y,z}$, Eq.~\ref{eq:c_theta} can be inverted as
\begin{equation}\label{eq:n_arccos}
	n_{x,y,z} = \cos \Theta_{x,y,z}=   1 + \frac{4(Q_o - \bar{c}_{x,y,z})}{ Q_a} ,
\end{equation}
for $|1 + 4 (Q_o - \bar{c}_{x,y,z})/Q_a|\leq 1$, so that $\Theta_{x,y,z}$ is real. 
To gain any useful information from the SWAP test, the attacker still needs to know $Q_o$ and $Q_a$ (Eq.~\ref{eq:c_theta}).
This provides an additional layer of security to the vault protocol, since now the attacker also needs to consume some of the tokens in the bill to obtain these values.
In the following, we conservatively assume that the attacker has perfect knowledge of these quality parameters.

The attacker can subdivide the $N$ qubits of the token into three equal parts of $N/3=30$ qubits and perform measurements along the three axes to gain more information about $\ket{\psi}$, as illustrated in Fig.~\ref{fig:attack_bloch}.
Now the question is how the attacker can optimally estimate $\ket{\psi}$ based on Eq.~\ref{eq:n_arccos} and the measured values of $\bar{c}_{x,y,z}$.
The most straightforward method is to perform a direct inversion tomography $\hat{n} = (n_x, n_y, n_z) / |n|$, but this was demonstrated to be less efficient compared to other estimation methods~\cite{qtoken_2}.
In this work, we employ an improved version of the maximum likelihood estimation (MLE) method as reported in Ref.~\cite{qtoken_2}, where the three values of $n_{x,y,z}$ constrained to $|\hat{n}|=1$ are simultaneously optimized by numerically maximizing the joint binomial probability for the measurements, instead of analytically solving an overdetermined set of equations.
More details on the mathematical implementation of the MLE are given in Sec.~\ref{sec:methods}.

After the probing of the genuine tokens, the attacker uses the estimated angles to prepare fake tokens and attempt authentication with the verifier, who accepts them if they are measured below the SWAP test acceptance threshold $c^*$.
This attack scenario was implemented in the three IBMQ considering 400 token states uniformly distributed over the Bloch sphere with optimal $N=90$.
The resulting acceptance probabilities for fake tokens $p_f[\bar{c} \leq c^*]$ as a function of the quality factors $Q_a$ are shown in Fig.~\ref{fig:attack_benchmark} and in Tab.~\ref{tab:params}, while the distributions of the SWAP test observable $\bar{c}$ from forged tokens are given in the GitHub repository~\cite{git_repo}.
The three processors present $p_f$ probabilities below 0.8, while the probability for accepting genuine tokens is $p_b \geq 0.99$ by construction (Eq.~\ref{eq:pb_c*}).
The $p_f$ values are within the same range of uncertainty, obtained by the bootstrap method~\cite{bootstrap}.
Despite that, as shown by the guide line obtained with a linear fit, these values agree with the trend observed in Refs.~\cite{qtoken_1, qtoken_2}, where better quality parameters lead to increased security of the tokens.

The probabilities for accepting single genuine $p_b$ and fake ones $p_f$ are relatively close.
However, as discussed in Sec.~\ref{sec:protocol}, the quantum bill security stems from the multiple tokens which it contains.
The probability then of a Type 2 error where a fake bill with $M$ tokens is accepted $\mathcal{P}_f$ follows a cumulative binomial distribution, as in Eq.~\ref{eq:Pb}, but substituting $p_f$ for the bank value $p_b$.
Fig.~\ref{fig:attack_benchmark}~(b) shows these probabilities of acceptance from the bank $\mathcal{P}_b$ and forger $\mathcal{P}_f$ as a function of the ratio $m^*/M$, for $M=8$ and $M=100$ in IBMQ Kingston.
As the ratio $m^*/M$ increases, the probabilities of acceptance $\mathcal{P}$ decrease, resulting from the sum of binomial distributions.
However, the decrease in $\mathcal{P}_f$ starts at lower $m^*/M$ values than $\mathcal{P}_b$, given that $p_f<p_b$.
For the larger $M$, Eq.~\ref{eq:Pb} approaches a rectangular function, making the protocol safer, since the acceptance threshold $m^*$ can be chosen such that $\mathcal{P}_b$ is still large but $\mathcal{P}_f$ is already decayed.

\begin{table*}[t!]
	\centering
	\setlength{\tabcolsep}{10pt}
	\renewcommand{\arraystretch}{1.3}
	\begin{tabular}{|c||c|c|c|c|c|}
		\hline
		IBMQ & $Q_a$ & $Q_o$ & $c^*(N=90)$ & $p_f$ & $1\sigma$ of $\mathcal{P}_f(M=200)$ \\ \hline
		Kingston & 0.892 (0.018) & 0.029 (0.004) & 0.078 (0.005) & 0.66 (0.06) & $10^{-33}$ -- $10^{-21}$ \\ 
		Fez & 0.818 (0.018) & 0.041 (0.004) & 0.100 (0.007) & 0.70 (0.05) & $10^{-25}$ -- $10^{-15}$ \\
		Marrakesh & 0.776 (0.020) & 0.084 (0.005) & 0.145 (0.009) & 0.75 (0.04) & $10^{-18}$ -- $10^{-10}$ \\ \hline
	\end{tabular}
	\caption[Quality Parameters from IBMQ Hardware]{
		Quality Parameters from IBMQ Hardware, as defined in Tab.~\ref{tab:variables}.
		Uncertainties in $Q_a$ and $Q_o$ are obtained from the standard deviation of the fitted parameters, while in $c^*$ and $p_f$ they are obtained from bootstrap method~\cite{bootstrap}.
		$1\sigma$ confidence intervals of $\mathcal{P}_f(M=200)$ are calculated from the propagated error of $p_f$ and $m^*$ and with the acceptance threshold $m^*$ set such that $\mathcal{P}_b \geq 1-10^{-4}$.
		Kingston presents better quality parameters and an improved resulting safety, while Marrakesh is the worst performing hardware.
	}
	\label{tab:params}
\end{table*}

To further characterize this effect, Fig.~\ref{fig:attack_benchmark}~(c) shows the values of $\mathcal{P}_f$ in all three IBMQ, with $m^*$ fixed such that the probability to accept genuine bills is $\mathcal{P}_b \geq 1-10^{-4}$.
Likewise, the probability for successful forgery $\mathcal{P}_f$ decays in an exponential regime with $M$.
As in Fig.~\ref{fig:benchmark}~(c), the decay is not monotonic due to the discretization noise of the variables.
Tab.~\ref{tab:params} provides the 1$\sigma$ confidence interval for $\mathcal{P}_f$ at $M=200$ with propagated errors from the $p_f$ values and $m^*$, leading to a wide range for $\mathcal{P}_f$.
Nevertheless, values of $\mathcal{P}_f<10^{-10}$ are observed in the worst performing hardware, assuming the pessimistic upper bound of the uncertainty.
Another important observation is the contrast in the security of the different hardware, where small improvements in the quality parameters can yield significant gains in security.
This shows a potential improvement to the protocol with the ongoing quantum hardware evolution~\cite{qcomputers_evolution, grand_challenge, token_fab_1, token_fab_2}.
But still, compared to the results reported in Ref.~\cite{qtoken_1} for an ensemble-based quantum token protocol with classical state information, these security metrics show that the elimination of the classical state information demands an increase in $M$ to compensate for the loss of efficiency in the authentication method.

\section{Discussion \& Outlook}\label{sec:discussion}

Quantum tokens envision to accomplish fundamentally unclonable authentication schemes, with broad applications to quantum money and network applications.
However, an overlooked vulnerability is an attack on the classical side-information regarding the quantum states of these tokens.
In this work, we answer this question with a conceptual shift from current quantum token proposals by symmetrically protecting both sides (issuing agent and user) with quantum no-cloning.
In the quantum vault proposal, the classical information $(\theta, \phi)$ about the user's token is eliminated by instead preparing two pairs of tokens with $N$ qubits in the same state $\ket{\psi}^{\otimes N}$, one given to the user and the other kept at the bank's vault.
The authentication is then performed through a SWAP test between the qubits in the tokens from the vault and the user.

We benchmark the whole protocol in a hardware-agnostic framework employing three IBMQ, with minimal assumptions about the hardware.
Besides paving the way towards applications with different platforms, the intra-chip results from the IBMQ provide an upper bound on the performance of concrete instantiations of the method in quantum networks, where additional errors are present, such as photon losses during token transfer or memory decoherence during storage.
Based on the definition of the $\bar{c}$ observable obtained from the SWAP tests, we are able to identify quality parameters (Tab.~\ref{tab:params}) for the hardware and to establish a criterion for acceptance. 
We further consider a viable estimate-and-forge attack to either one of the two tokens, where the attacker performs SWAP tests to gain information about the tokens and then use this information to create fake ones.
Considering single tokens, an attacker has a high probability of success, demonstrating the efficiency of the attack method.
However, in quantum bills composed of $M=200$ tokens (Eq.~\ref{eq:Psi}), we obtain probabilities of successful attacks $\mathcal{P}_f<10^{-10}$ even in the worst performing processor and most conservative uncertainty, while the probability for genuine tokens to be accepted is $\mathcal{P}_b \geq 1-10^{-4}$.
By comparing the security metrics with previously reported results for an ensemble-based quantum token protocol with classical state information in IBMQ processors~\cite{qtoken_1}, we observe that the elimination of the classical state information lowers the authentication efficiency of genuine tokens, thus requiring more tokens per bill.

In practical terms, the results indicate that trusted or un-trusted parties can have access to the authentication interface of tokens in the vault, without being able to impersonate the users and compromise security.
In this sense, the keys can be `semi-public', but still a centralized institution such as a bank needs to store and make them accessible.
These results further support previous proposals for using SWAP tests in quantum token verification~\cite{qtoken_network_authentication, public_qcoins}.
The quantum vault scheme as presented here is intrinsically non-anonymous, since the bank needs to know who is the owner of the token in order to perform the comparison.
This confers the governance properties of traceability and revocability~\cite{anonymous_qtoken, anonymous_qtoken_2, qmoney_broken_2}, meaning that responsible authorities can trace a bill and revoke its use at any time, since the bank holds the vault copy.
Another practical consideration for the protocol is that the authentication consumes the quantum tokens, which implies that the user should receive several units of them at once by the bank~\cite{public_qcoins}, or have them replenished through a quantum channel.
With ongoing hardware evolution~\cite{qcomputers_evolution, grand_challenge, token_fab_1, token_fab_2}, a further improvement of the method can be obtained with optimized ensemble states comparison, as proposed by M. Fanizza et al. (2020)~\cite{optimal_swap}.

The main technological limitation for the commercial application of such scheme lies in the short lived quantum memories presently available.
Even though coherence times of several hours have been observed in nuclear spins in noble gases~\cite{noble_gas_T2_1, noble_gas_T2_2} that can even be made mobile, these quantum systems still lack a reliable interface with flying qubits.
Other trapped atomic and molecular platforms~\cite{trapped_ion, trapped_neutral} can have a viable interface with flying qubits, long coherence, high fidelity gates, but can hardly be made mobile due to the required ultra-high vacuum or low temperatures.
The same limitations apply to superconducting qubits, which provide an efficient platform for validating this protocol, but require mK temperatures for operation.
Finally, a class of quantum systems that has great potential as quantum token platforms is the color center defects in solids~\cite{CC1, CC2, teleport2}.
With ongoing evolution of their fabrication techniques~\cite{grand_challenge,token_fab_1, token_fab_2, qtoken_tim}, these systems exhibit naturally long coherence times at room temperature and pressure, with strong optical interface~\cite{qtoken_tim}.
Another interesting feature of these systems is the ensemble inseparability guaranteed by diffraction limited optical initialization and readout~\cite{optical_pumping, teleport2}, which limits the attacker to perform SWAP tests along a single axis, but would also require a reformulation of the whole SWAP test authentication not based on Bernoulli trials.
Alternative approaches that circumvent the necessity for long lived quantum memories are the S-money proposals, which instead support their security on relativistic principles~\cite{smoney_1, smoney_2}.

Even though the estimate-and-forge attack model considered here covers a wide range of possible forms of attack, it does not configure a full security proof.
For instance, better cloning efficiency can be obtained if the attacker is not restricted to performing SWAP tests, but instead can also perform usual projective measurements.
This restriction could be enforced in some architectures by a hardware fabrication which restricts arbitrary gate operations to the token qubits and enable measurements only on the auxiliary qubit.
An alternative type of attack are substitution attacks~\cite{qtoken_attack}, where the attacker simply swaps the genuine tokens, without violating the no-cloning theorem.
This attack form can also be avoided with hardware constraints on arbitrary SWAP gates to the token qubits.
In addition, exact characteristics of each physical platform can be exploited by side-channel attacks~\cite{side_channel_attack, qtoken_1, bsi_qkd_attacks} or better Bayesian estimation methods with different budget allocation of the ensemble could potentially be used by the attacker~\cite{qtoken_2}.
As argued by A. Behera and O. Sattath (2024)~\cite{public_qcoins}, many theoretically proposed constructions for public quantum money were later broken, while current ones are also susceptible to being broken~\cite{qmoney_broken_1, qmoney_broken_2}.
This motivates further theoretical examination of the quantum vault proposal and highlights the hardware-agnostic approach developed here of employing cloud-based quantum processors as a viable and practical means for testing quantum money and communication methods.

\section{Methods}\label{sec:methods}

\subsection*{IBMQ Implementation}

The quantum vault was implemented using Qiskit SDK version 2.3.0~\cite{qiskit} and Qiskit IBMQ Runtime version 0.45.1, as open-source provided at the author's GitHub repository~\cite{git_repo}.
The three processors used in this work are from the Heron r2 family with similar architecture~\cite{qcomputers_evolution}, having the same native operations and 156 qubit count.
Given that their parameters exhibited fluctuations in time which affect these results, all measurements were carried out within a few hours on 10.09.2026, with the calibrations reports provided by IBMQ also given in the repository~\cite{git_repo}.
The high-level representations of the quantum circuits were transpiled to the native gates of the hardware without optimization or gate approximation, in order to keep the circuit as close as possible from the model.
The gates are translated by synthesis to the densest layout qubits, with longest coherence, shortest gate times and errors.
In Kingston, qubits 10 and 11 were used for the vault with 12 as the auxiliary.
In Fez, 109 and 110 are the vault, 111 is the auxiliary.
Lastly, in Marrakesh, 135 and 139 are the vault, 155 is the auxiliary.
All circuits were run in the sampler mode.

\subsection*{Maximum Likelihood Estimation}

As discussed in Sec.~\ref{sec:security}, the attacker probes the $N$ genuine tokens in the state $\ket{\psi}^{\otimes N}$ along the vector $\hat{n}=(n_x, n_y, n_z)= (\sin\theta \cos\phi, \sin\theta \sin\phi, \cos\theta)$ by subdividing them in three parts and performing measurements along the three axes, thus obtaining $\bar{c}_{x,y,z}$.
Following Eq.~\ref{eq:c_theta}, the model's probability to measure $\bar{c}_i=1$ on axis $i$ as a function of the trial parameters $(\theta, \phi)$ is
\begin{equation*}
	P_i(\theta, \phi) = Q_{o} + \frac{Q_{a}}{4} \left[1 -  n_i \right].
\end{equation*}
Since the individual $c_n$ measurements are Bernoulli trials of binary outcomes and each repetition is independent, the likelihood $\mathcal{L}(\theta, \phi | \bar{c}_{x,y,z})$ for measuring $\bar{c}_{x,y,z}$ is given by the product of the three binomial probabilities
\begin{multline*}
	\mathcal{L}(\theta, \phi | \bar{c}_{x,y,z}) = \\ \prod_{i \in \{x,y,z\} } \binom{N/3}{N\bar{c}_i/3} P_i^{N\bar{c}_i/3} (1- P_i)^{N(1-\bar{c}_i)/3}.
\end{multline*}
This is the function the attacker wants to maximize in order to estimate the state $\ket{\psi}$.
However, maximizing $\mathcal{L}(\theta, \phi |\bar{c}_{x,y,z})$ corresponds to minimizing
\begin{multline*}
	- \log \mathcal{L}(\theta, \phi | \bar{c}_{x,y,z}) = \\ -\sum_{i \in \{x,y,z\} } \frac{N}{3} \left[ \bar{c}_i \log (P_i) + (1-\bar{c}_i) \log(1-P_i) \right] + \textrm{const} ,
\end{multline*}
given that log is a monotonic function.
We can further drop the $N/3$ factor and the constant term, since they do not affect the minimization.
Finally, using the Nelder-Mead method~\cite{nelder_mead}, the optimal solution of $(\theta, \phi)$ is numerically found, with the direct inversion tomography as initial guess.
The Python implementation of the method can also be found at the GitHub repository~\cite{git_repo}.
Likewise, this method represents an improvement from the MLE reported in Ref.~\cite{qtoken_2}, by fully exploiting the information gained from the $N$ measurements without equation over-determination.

\section*{Data \& Code Availability}

All data used in this work and the code for implementing the quantum vault protocol and analyzing the results are open source under GPL3 license, provided at the author's GitHub repository~\cite{git_repo}.

\section*{Author Contributions}

L.T. conceptualized this work, developed the theory, implemented the protocols with IBMQ and analyzed the data.
B.N. acquired funding and supervised the work.
Both authors contributed to the writing and revision of the text.

\section*{Acknowledgments}

I (L.T.) am extremely grateful to the anonymous scientist who attended my poster presentation at the KIT Graduate School of Quantum Matter (KSQM) in 2023 and through a constructive discussion, gave me the seed for the idea that motivated this work.
This work was supported by the \textit{Bundesministerium für Forschung, Technologie und Raumfahrt} (BMFTR) under the project \textit{Diamant-basierte Quantenmaterialien} (DIAQUAM - n\textsuperscript{o} 13N16956) and by the Helmholtz Quantum Use Challenge project QT-Batt. 
We acknowledge the use of IBM Quantum services for this work. The views expressed are those of the authors and do not reflect the official policy or position of IBM or the IBM Quantum team.

\begin{table*}[hbt!]
	\centering
	\begin{tabular}{|c c|}
		\hline
		Variable & Physical Meaning \\  \hline
		$\theta$ & polar angle on the Bloch sphere \\
		$\phi$ & azimuthal angle on the Bloch sphere \\
		$\hat{R}(\theta, \phi)$ & rotation operator with angles $(\theta, \phi)$ \\
		$\ket{\psi}$ & token state with angles $(\theta, \phi)$ \\
		$\Theta$ & angle between two states with unit vectors $\hat{n}$ and $\hat{n}^\prime$: $\Theta = \arccos (\hat{n} \cdot \hat{n}^\prime)$ \\
		$c_n$ & $n$-th classical register for the SWAP test \\
		$N$ & number of qubits in the token \\
		$\bar{c}$ & SWAP test observable given by the average $\sum_{n=1}^{N} c_n/N$ \\
		$c^*$ & acceptance threshold of the SWAP test \\
		$Q_a$ & amplitude quality factor \\
		$Q_o$ & offset quality factor \\
		$p_b$ & probability that a genuine token is accepted \\
		$p_f$ & probability that fake token is accepted \\
		$M$ & total number of tokens in quantum bill \\
		$m^*$ & acceptance threshold for the minimum number of accepted tokens in a bill \\
		$\ket{\mathbf{\Psi}}$ & quantum bill state with $M$ tokens each with $N$ qubits $\bigotimes^M_{m=1} \ket{\psi}_m^{\otimes N}$\\
		$\mathcal{P}_b$ & probability that a genuine quantum bill is accepted \\
		$\mathcal{P}_f$ & probability that a fake quantum bill is accepted \\
		$\mathcal{L}(\theta, \phi | \bar{c}_{x,y,z})$ & likelihood for an attacker to measure $\bar{c}_{x,y,z}$ over the three axes \\ \hline
	\end{tabular}
	\caption[Glossary of Variables]{Glossary of variables.}
	\label{tab:variables}
\end{table*}

%%%%%%%%%%%%%%%%%%%%%%%%%%%%%%%%%%%%%%%%%%%%%%%%%%%%%%%
\pagebreak

\end{document}